# Dislocation-mediated plasticity in the Al$_2$Cu θ-phase

D. Andre[a,*], Z. Xie[a], F. Ott[a], J. T. Pürstl[b], N. Lohrey[a], W.J. Clegg[b], S. Sandlöbes-Haut[a], S. Korte-Kerzel[a]

[a]Institute of Physical Metallurgy and Materials Physics, RWTH Aachen University,

[b]Department of Materials Science and Metallurgy, University of Cambridge

* andre@imm.rwth-aachen.de (corresponding author)

## Abstract
The deformation behaviour of the intermetallic Al$_2$Cu θ-phase was investigated using atomistic simulations and micropillar compression, where slip on the unexpected {211} and {022} slip planes was revealed. Additionally, all possible slip systems for the intermetallic phases were further evaluated and a preference for the activation of slip systems based on their effective interplanar distances as well as the effective Burgers vector is proposed. The effective interplanar distance corresponds to the manually determined interplanar distance, whereas the effective Burgers vector takes a potential dislocation dissociation into account. This new order is: {211}½<111>, {022}½<111> and {022}<100>, {110}<001>, {310}<001>, {022}<011>, {110}½<111>, {112}<110> and {112}½<111> from high to low ratio of $d_{eff}/b_{eff}$. Also, data on the critical resolved shear stresses of several of these slip systems were measured.

Keywords: micropillar compression, CRSS, dislocations, plasticity, generalized stacking fault energy

## 1  Introduction

Composite materials containing metallic and intermetallic phases are nowadays widely used for advanced applications such as Ni-based superalloys for high temperature applications [1, 2]. Even though the metallic-intermetallic composite alloy Al-Al$_2$Cu is one of the most intensively studied eutectic composite materials in the literature [3-13], the plastic deformation behaviour of the intermetallic Al$_2$Cu θ-phase is not completely understood yet. So far, investigations of its plastic deformation behaviour at ambient temperatures were mostly hindered due to its brittle failure at low homologous temperatures. Consequently, investigations on the active glide systems were mainly performed at elevated temperatures on macroscopically scaled specimens. Kirsten [14] was the first who reported plasticity in the Al$_2$Cu θ-phase. According to their observations, slip occurred on three planes, namely the (111), (110) and (101) slip plane. Some of these slip planes, specifically the {111} and {101} planes, however, were already doubted by Ignat et al. [15] some years later, as Ignat et al. considered {111} planes and {101} planes as planes of low packing density, not taking into account the complexity of the ordered intermetallic phase. They claimed to observe slip on the {110} plane in either <001> or <111> direction, and reported further the activation of the {200}<001> and {112}<111> slip systems. This assumption was further supported by the results of Hamar et al. [16] who reported cleavage on {110} and {112} planes and Riquet et al.

[17] who reported that {110} planes form growth facets in the intermetallic dendrites. A later publication by Ignat et al. [18] focused on the deformation of Al-Al$_2$Cu eutectic samples at elevated temperatures revealing dislocations with Burgers vectors in [$\bar{1}$00] direction on (0$\bar{1}\bar{1}$), (00$\bar{1}$) and (0$\bar{1}$0) planes. It was further assumed that these straight dislocations are screw dislocations.

In addition to these experimental studies, theoretical approaches were used to gain insights into the dislocation motion of the Al$_2$Cu θ-phase. Recently, Zhou et al. [19] performed atomistic studies on the dislocation motion in the Al$_2$Cu θ-phase revealing glissile dislocations with a Burgers vector of <001> on (110), (010) and (310) planes at room temperature and moderately elevated temperatures. Furthermore, they deduced climb and collective glide of <001> dislocations resulting in a dislocation core change and an atomic shuffle in a later study [20]. Additionally, Liu et al. [9] reported the activation of the "unusual" {011} slip plane in the Al-Al$_2$Cu eutectic when assuming an orientation relationship of (110)$_{Al2Cu}$||(111)$_{Al}$ using atomistic simulations. They concluded that slip transmission may be the reason for the activation, as Zhou et al. [19] did not observe this slip system using atomistic simulations on the pure intermetallic phase.

Nowadays, the challenge to study plastic deformation of brittle materials at ambient temperatures can be met by the use of new experimental techniques, such as micropillar compression. This method was developed about 16 years ago with the intention to study the effect of size on mechanical properties and initially applied predominately to face centred cubic metals [21]. Later, it was found that the method also allows suppression of cracking during deformation of brittle materials due to the reduced sample size [22-25].

In this work, we therefore performed micropillar compression experiments to investigate the plastic deformation behaviour of the intermetallic Al$_2$Cu θ-phase and used atomistic simulations to achieve further insights into the underlying slip systems, in particular the active slip planes and potential Burgers vectors. We also performed an in-depth analysis of the spacings of the underlying potential slip planes in the Al$_2$Cu unit cell. The idea behind this approach was two-fold: (i) to identify the likely planes of interest in order to inform and guide the computationally simulations and also (ii) to provide a set of fundamental measures (the ratios of interplanar spacing to Burgers vector) that act as a basis to assess the validity of conventionally expected correlations between the crystal structure and the observed critical stresses for dislocation motion against the resistance of the crystal lattice in this intermetallic phase.

## 2 Experimental procedure

### 2.1 Single crystal microcompression experiments

Al$_2$Cu samples were cast using a vacuum induction melting furnace (MAM1 Edmund Bühler GmbH). Metallographic sample preparation was done using grinding and polishing. Electron backscatter diffraction (EBSD) analysis was performed to analyse the crystallographic orientations of the investigated grains. Micropillars with a circular cross-section, a diameter of approximately 2 µm and an aspect ratio of 2-2.5 were milled using a focussed ion beam (FIB, Helios Nanolab 600i, FEI). The resulting average taper amounted to about 3.5°. Micropillar compression was performed using an iNano nanoindenter (Nanomechanics, Inc.) with a diamond flat punch at a loading rate of 0.5 mN/s until strain bursts occurred. The resulting load-displacement response was corrected for the Sneddon sink-in according to Frick et al.

[26] and converted into engineering stress and strain values, which are based on the initial pillar diameter (measured at the top of the pillar) and length, respectively. Images of the pre- and post-deformed pillars were taken under a tilt angle of 45° using an acceleration voltage of 10 kV. The tilt angles of the slip steps formed at the pillar surface were correlated to the tilt angles of potential glide planes of the corresponding crystal system using VESTA (Visualization for Electronic and Structural Analysis, [27]) taking into account the closest packed planes and shortest directions of the crystal structure according to Ignat et al. [15]. OIM analysis™ (EDAX Inc.) was used for Schmid factor calculations of the identified slip systems. The critical resolved shear stresses (CRSS) were calculated from the engineering stress at the onset of yielding, corresponding to the first strain burst, and the Schmid factor of the slip system as identified by scanning electron microscopy (SEM) and EBSD.

## 2.2 Simulation set-up

Molecular statics (MS) simulations were performed on the $Al_2Cu$ θ-phase to calculate generalized stacking fault energy surfaces, known as γ-surfaces, of potential slip planes. The interatomic interactions were modelled by an embedded-atom-method (EAM) potential by Liu et al. [28] and an angular-dependent potential (ADP) by Apostol and Mishin [29] for the Al-Cu system. The simulations were carried out using the molecular dynamics (MD) software package LAMMPS [30].

The simulation samples contain more than 4000 atoms. The cell sizes within the studied slip plane $l_{x,y}$ are larger than $2 \times$ cutoff radius ($r_{cut}$) of the interatomic potential. The cell size in z-direction $l_z$ is about $5 \times l_{x,y}$. Periodic boundary conditions (PBC) were applied along the x- and y-axes parallel to the slip plane. Non-periodic boundary conditions (nPBC) were used along the z-direction perpendicular to the slip plane.

The γ-surface was determined by incrementally shifting one half of the crystal along the corresponding slip direction across the slip plane with an incremental displacement of $\vec{b}/20$. After each displacement step, the crystal was allowed to relax in the z-direction perpendicular to the slip plane using the FIRE algorithm [31, 32] until the force-norm was below $10^{-8}$ eV/Å. The γ was calculated according to Equation 1:

$$\gamma = \frac{E - E_0}{A}, \qquad \text{Eq. 1}$$

where the energy difference between the shifted crystal and the initial one is $E - E_0$, and $A$ is the area of the studied slip plane $A = l_x l_y$.

## 3 Results

In order to gain a deeper understanding of the mechanisms governing plasticity in the $Al_2Cu$-phase, we performed micropillar compression experiments on differently oriented pillars. This allows us to unravel the active slip planes and estimate slip directions (Section 3.1). A closer look at the geometric structure of the corresponding planes and the ones considered in literature was taken in order to identify the planes of interest including their effective interplanar distances (Section 3.2). The additional use of MS simulations on the potential slip planes then enabled us to obtain information on the generalised stacking fault energies and Burgers vectors (Section 3.3) – with all results taken together giving an in-depth view into the potential plasticity mechanisms involving dislocation slip in Al$_2$Cu (Section 4).

## 3.1 Micropillar compression experiments

The deformation of 15 micropillars (MP) with different initial crystal orientations revealed slip on three primary slip systems and two secondary slip systems. All observed slip systems including their Schmid factors and critical resolved shear stresses are summarized in Table 1.

Table 1: Slip systems including their compression axis, Schmid factor, m$_S$, yield stresses, σ$_y$, critical resolved shear stresses (CRSS) and average CRSS per slip system for the Al$_2$Cu θ-phase.

| Pillar | Slip system | Pillar axis | m$_S$ | σ$_y$ [GPa] | CRSS [GPa] | Av. CRSS [GPa] |
|---|---|---|---|---|---|---|
| 1-3 | (211)½[1$\bar{1}\bar{1}$] ($\bar{1}$10)½[$\bar{1}\bar{1}\bar{1}$] | (1 20 10) | 0.47 0.48 | 1.61 ± 0.11 - | 0.76 ± 0.05 - | 0.76 ± 0.05 - |
| 4 | (02$\bar{2}$)½[$\bar{1}\bar{1}\bar{1}$] (121) ½[$\bar{1}$1$\bar{1}$] | (18 1 21) | 0.48 0.48 | 1.71 - | 0.82 - | 0.88 ± 0.07 |
| 5,6 | (02$\bar{2}$)½[$\bar{1}\bar{1}\bar{1}$] | (10 3 14) | 0.49 | 1.67 ± 0.18 | 0.82 ± 0.09 | |
| 7 | (202)½[11$\bar{1}$] | (1 $\bar{5}$ 8) | 0.5 | 1.87 | 0.94 | |
| 8,9 | (202)½[11$\bar{1}$] | ($\bar{7}$ $\bar{5}$ 17) | 0.45 | 2.06 ± 0.03 | 0.93 ± 0.02 | |
| 10,12 | (20$\bar{2}$)[0$\bar{1}$0] | ($\bar{18}$ 21 16) | 0.49 | 2.10 ± 0.32 | 1.03 ± 0.16 | 0.97 ± 0.12 |
| 11 | (022)[100] | ($\bar{18}$ 21 16) | 0.49 | 2.19 | 1.07 | |
| 13 | (202)[010] | (14 $\bar{17}$ 10) | 0.49 | 1.79 | 0.87 | |
| 14,15 | ($\bar{2}$02)[0$\bar{1}$0] | ($\bar{14}$ 21 13) | 0.5 | 1.83 ± 0.16 | 0.91 ± 0.09 | |

For pillars 1-3, (211)½[1$\bar{1}\bar{1}$] was found as primary slip system and ($\bar{1}$10)½[$\bar{1}\bar{1}\bar{1}$] as secondary slip system. The differentiation between primary and secondary slip system is based on the post-compression images. The orientation of the unit cell including its activated slip systems (front view and view on pillar top) and the corresponding stress-strain response is given in Figure 1 d) and e). The first strain burst is generally believed to be predominantly linked to the activation of the primary slip system. As it is not possible to correlate strain bursts to the secondary slip system, no CRSS value was calculated.

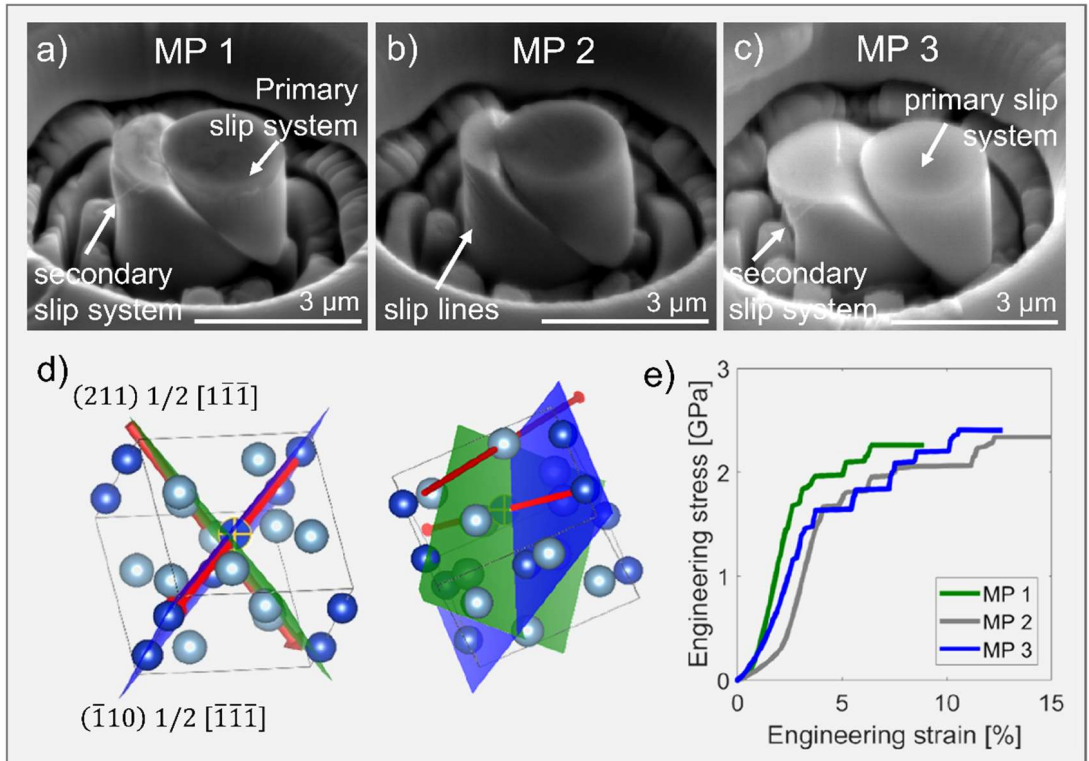

Figure 1: a-c) SE-images of single-crystalline Al$_2$Cu θ-phase micropillars 1-3 with compression direction || [1 20 10] under a tilt of 45°. Micropillar 1-3 in a)-c) reveal slip on the primary (211)½[1$\bar{1}\bar{1}$]-slip system and the secondary ($\bar{1}$10)½[$\bar{1}\bar{1}\bar{1}$] slip system (a) & c)). The Al$_2$Cu unit cell with view on pillar front and view on pillar top are shown in (d). The left unit cell represents the front view of the pillar, whereas the right unit cell corresponds to a view of the pillar top in order to define the Burgers vector. Further considerations on the {211} and {110} slip planes can be found in section 4. e) The resulting engineering stress strain response reveals strain bursts upon start of plastic deformation.

For six micropillars (MP 4-9) in Figure 2 a-g), the {022}½ < 111 > slip system was identified as primary slip system and {121}½ < $\bar{1}1\bar{1}$ > as secondary slip system.

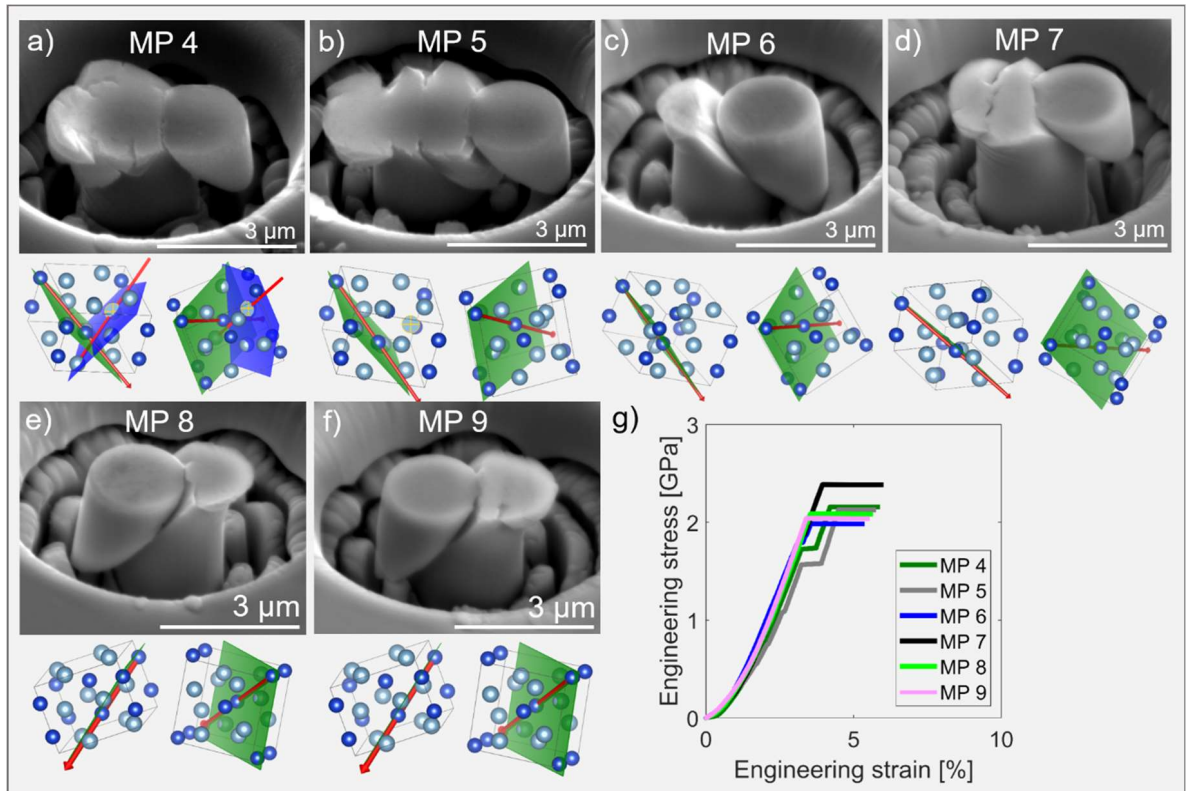

Figure 2: a-f) SE-images of micropillars 4-9 under a tilt of 45° including their slip systems in the $Al_2Cu$ unit cell. The first unit cell represents the view on the pillar front, whereas the second unit cell represents the view on the pillar top. Plasticity occurred on the $\{022\}½<111>$ primary slip system for all pillars (micropillars 4-9). In addition, for micropillar 4, the $(121)½[\bar{1}1\bar{1}]$-slip system was identified as secondary slip system. Further considerations on the $\{022\}$-slip plane can be found in section 4. The corresponding engineering stress-strain curves are depicted in g).

Slip on $\{022\}<010>$ occurred in six micropillars, namely MP 10-15 (Figure 3) at an average CRSS of 0.97 ± 0.12 GPa.

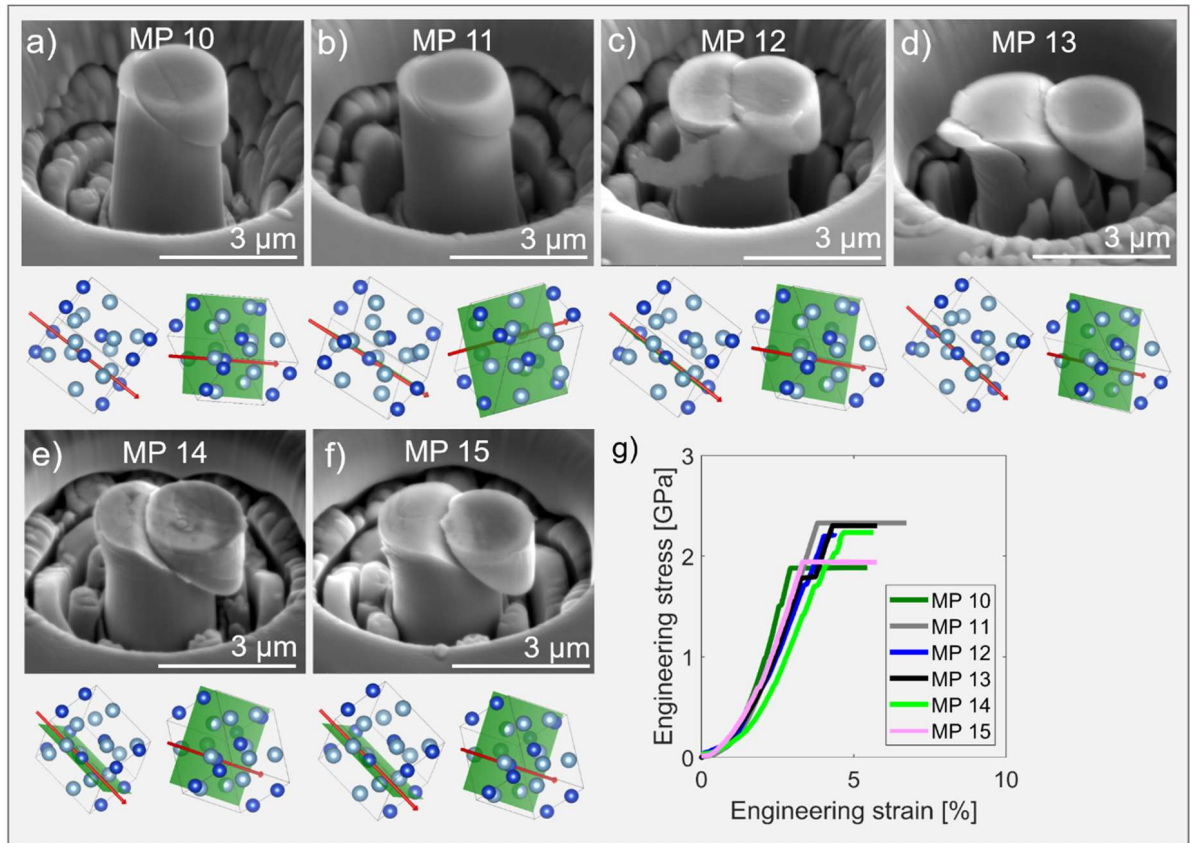

Figure 3: a-f) SE-image of pillars 10-15 under a tilt of 45° including their slip systems in the Al2Cu unit cell. The first unit cell represents the view of the pillar front, whereas the second unit cell represents the pillar top view. The pillars slipped on the $\{022\} <010>$ slip system. Further considerations on the $\{022\}$ slip plane can be found in section 4. The corresponding engineering stress-strain curves are depicted in g).

### 3.2 Effective interplanar spacing $d_{eff}$ of slip planes

Within this section, a closer look on the most important slip systems of the $Al_2Cu$ Θ-phase was taken and the effective interplanar distances $d_{eff}$ of the complex ordered unit cell were manually determined for lattice parameters of a = b = 6.06700 Å and c = 4.87700 Å according to Meetsma et al. [33]. For a better overview, all considered slip systems including their full Burgers vectors in the $Al_2Cu$ unit cell are given in Figure 4.

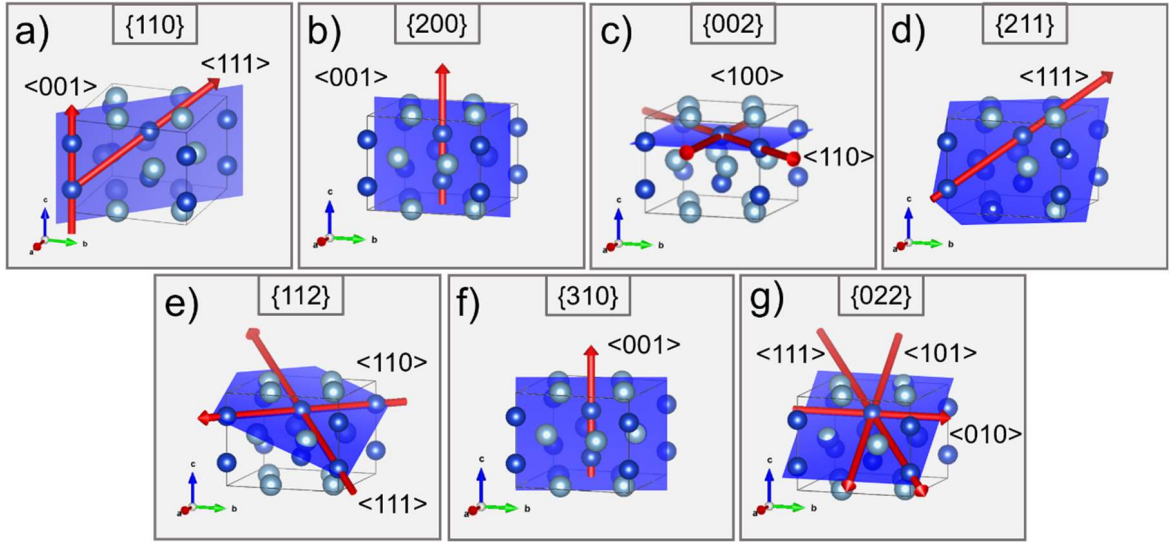

Figure 4: Overview of the most important slip systems of the Al$_2$Cu Θ-phase: a) representing the {110} plane with <111> and <001> directions, b) representing the {200} plane with <001> direction, c) representing the {002} plane with <100> and <110> directions, d) representing the {211} plane with <111> direction, e) representing the {112}-plane with <110>- and <111>-directions, f) representing the {310} plane with <001> direction, g) representing the {022} plane with <111>, <101> and <010> directions in the Al$_2$Cu unit cell.

Effective interplanar spacing of the {110} plane

The {110}½<111> and {110}<001> slip systems (Figure 5 a)) reveal a stacking sequence of Cu-Al$_1$-Al$_2$-Al$_1$-Cu-Al$_3$-Al$_4$-Al$_3$. The largest interplanar distance, also known as effective interplanar distance $d_{eff}$, of 1.3565 Å exists between layers Al$_1$ and Al$_2$ and layers Al$_3$ and Al$_4$, respectively. The shortest directions on this slip plane are ½<111> and <001> with a length of 4.931 Å and 4.877 Å, respectively, see Figure 5 b) and c).

Effective interplanar spacing of the {200} plane

The effective interplanar distance of the {200}<001> slip system is calculated to be 1.1151 Å (Figure 5 d)), corresponding to the distance between two Al-layers (Al$_1$-Al$_2$ and Al$_3$-Al$_4$). The full Burgers vector <001> has a length of 4.877 Å (Figure 5 e)).

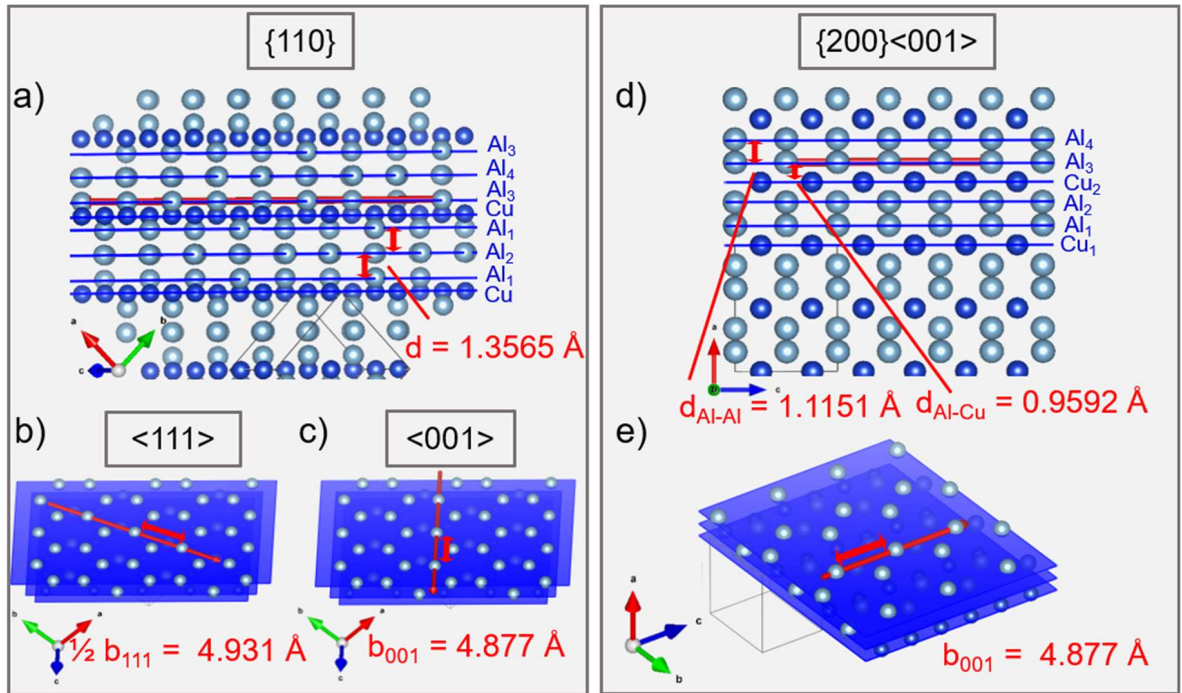

Figure 5: a) the Al$_2$Cu unit cell, bright blue balls represent Al- and dark blue balls represent Cu-atoms. The side view onto the closed packed {110}-planes reveals the atomic layers and the largest interplanar distance d is also indicated. b) view onto the {110}½<111> slip system, c) view onto the {110}<001> slip system, d) side view onto the {200}<001> slip system with the three different potential planes including their interplanar distances $d_{Al-Al}$ and $d_{Al-Cu}$, e) view onto the {200}<001> slip system. The coordinate system of the crystal structure is given in each plot for a better orientation.

Effective interplanar spacing of {002} plane

The {002}-slip planes (Figure 6 a)) are arranged as alternating layer of Al and Cu atoms with a stacking sequence of Cu-Al$_1$-Cu-Al$_2$ and an effective interplanar distance of 1.2193 Å. Further, potential slip directions are given in (Figure 6 b-c)).

Effective interplanar spacing of the {211} plane

The {211}½<111> slip system (Figure 6 d-e)) consists of a layered structure with each complex layer containing both atomic species (corresponding to 4 layers in Figure 6 d)). It is assumed that dislocation motion on the {211} plane can only occur between the sublayers, due to the high density of atoms within each layer and their unavoidable interference upon dislocation movement. These complex layer has an interplanar spacing of 2.3710 Å, when assuming that the middle of each plane defines the effective interplanar distance.

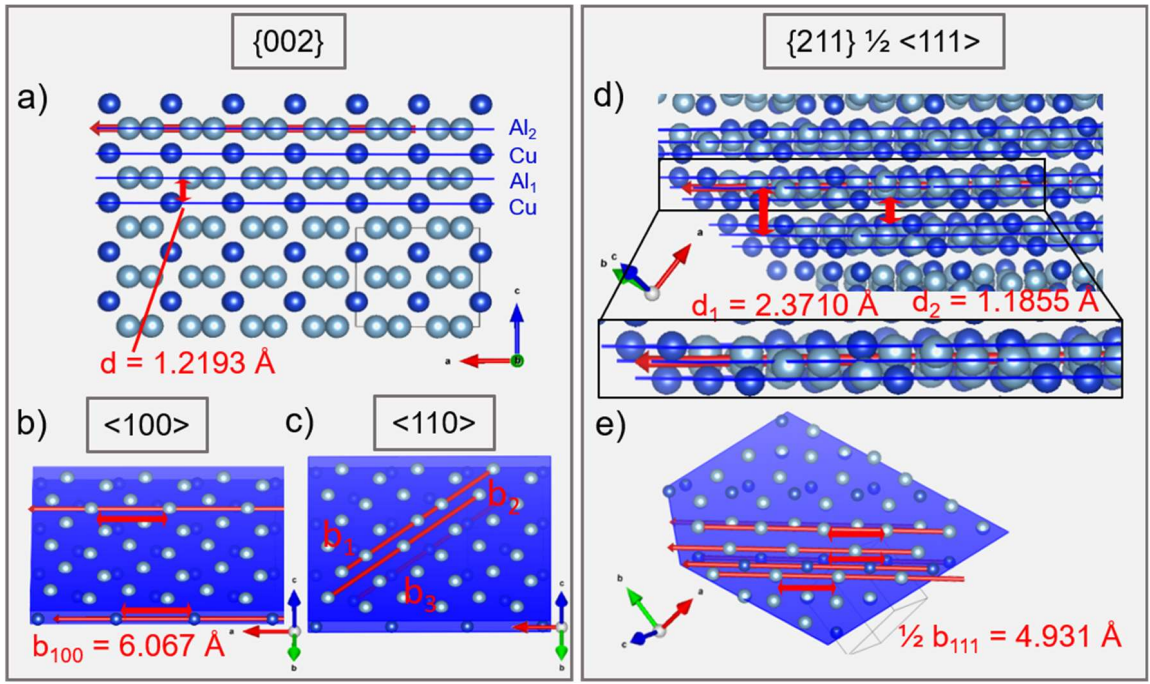

Figure 6: View onto the Al₂Cu unit cell with its possible slip systems with bright blue balls representing Al- and dark blue balls representing Cu-atoms: a) side view onto closed packed {002} planes and the interplanar distance d, b) view onto the {002}<100> slip system, with possible slip vectors in the Al- and Cu-planes c) view onto the {002}<110> slip system, with its three different slip options, $b_1$, $b_2$ and $b_3$, d) side view onto the {211}½<111> slip system, e) view onto the {211}½<111> system.

Effective interplanar spacing of the {112} plane

The {112}½<111> slip system is composed of a layered structure containing pure Al and Al-Cu layers (Figure 7 a)). The largest interplanar distance of 0.7793 Å exists between two Al-layers. Two representative Al-layer and shortest directions are represented in Figure 7 b-c).

Effective interplanar spacing of the {310} plane

The {310} planes have a complex layered structure with an interplanar spacing of 1.2133 Å (Figure 7 d)) when considering the distance between the closest Al-layers. However, assuming that each slip plane consists of both types of atoms (corresponding to 6 planes in Figure 7 d)), a dislocation must displace Al- and Cu-atoms simultaneously (Figure 7 e)) and the larger interplanar spacing of 1.9186 Å might represent the correct effective interplanar spacing.

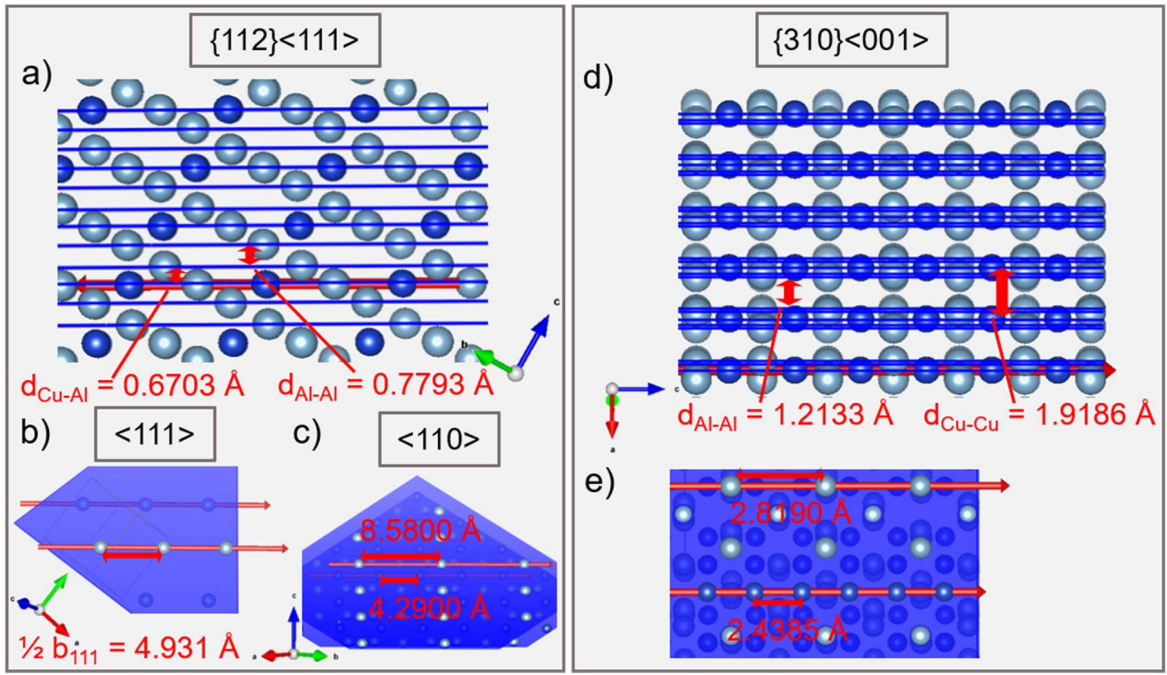

Figure 7: View onto the Al$_2$Cu unit cell with its possible slip systems with bright blue balls representing Al- and dark blue balls representing Cu-atoms: a) side view onto closed packed {112} planes and the two interplanar distances $d_{Cu-Al}$ and $d_{Al-Al}$, b) view onto the {112}½<111> slip system, with possible slip vector, c) view onto {310}<001> slip system with interplanar distance $d_{Al-Al}$, d) side view onto the {310}<001> slip system, with its different interatomic distances, e) view onto the {310}½<001> slip system, with its possible slip vector.

Effective interplanar spacing of the {022} plane

The {022} planes are arranged in complex layers with each layer containing Cu and Al-atoms, (see Figure 8). If the interplanar distance is measured assuming that Cu-atoms define the position of the plane, $d_{Cu-Cu}$ equals 1.9006 Å. When assuming that the lower and upper Al-planes define the effective interplanar distance $d_{Al-Al}$, it equals 1.2019 Å. However, as the atomic density in each of the Al$_2$Cu layers is high and dislocation motion in one of these Al$_2$Cu layers is therefore assumed to involve the entire plane to move, neglecting the possibility of a complex deformation process, e.g. synchroshear, it is further assumed that the larger interplanar spacing is correctly representing the effective interplanar spacing of this plane. Furthermore, the length of all possible full Burgers vectors within this plane were evaluated and are depicted in Figure 8 b-d).

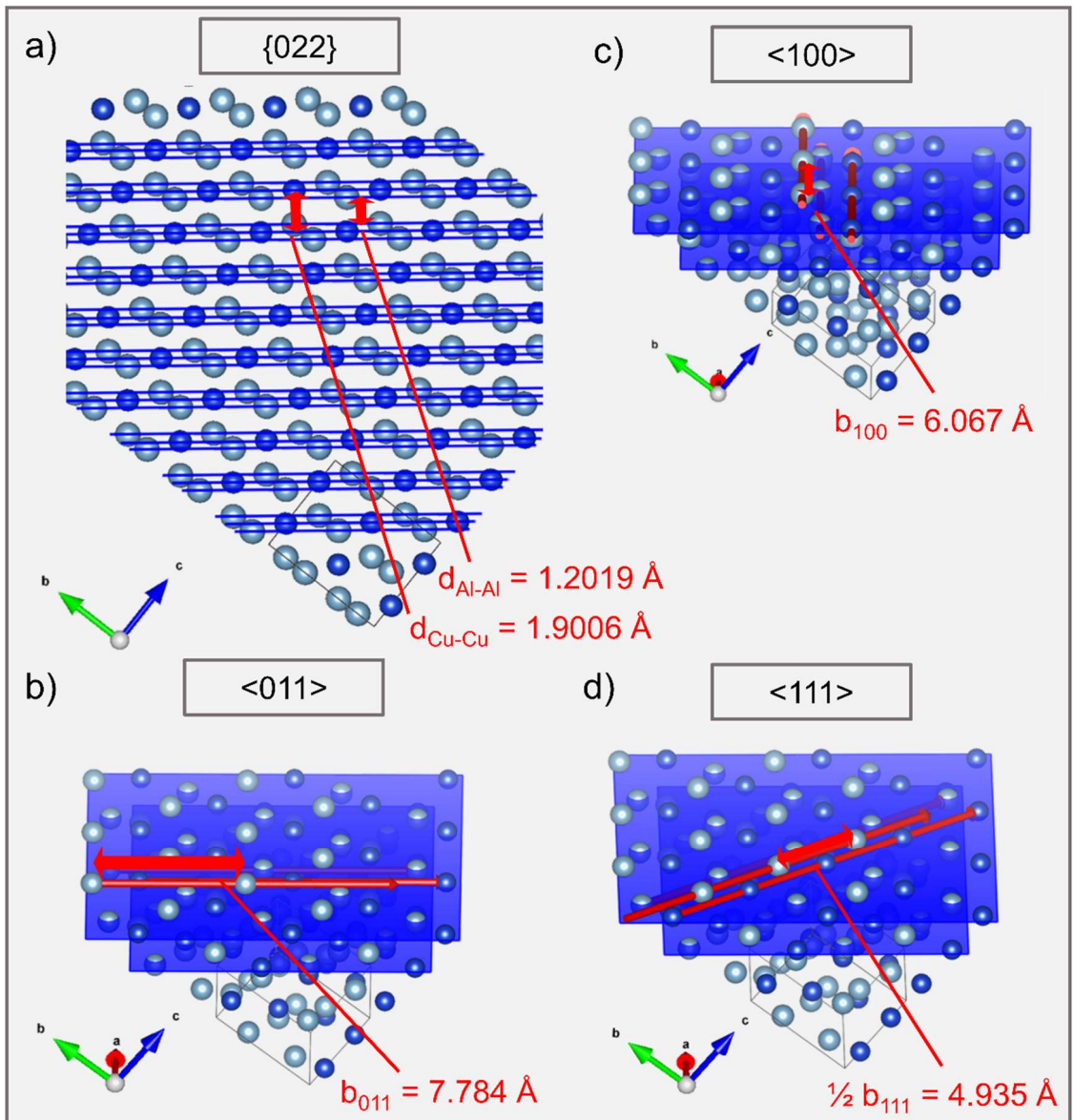

Figure 8: View onto the $Al_2Cu$ unit cell with its possible slip systems with bright blue balls representing Al- and dark blue balls representing Cu-atoms: a) side view onto the closed packed {022} slip system revealing the stacking sequence of Al- and Cu-planes and both options to measure the interplanar distance d, b) Burgers vector for slip in <011> direction, c) Burgers vector for slip in <100> direction, d) Burgers vector for slip in <111> direction.

### 3.3 Generalized stacking fault energy surfaces of potential slip systems

Atomistic simulations were performed in order to obtain and compare the generalized stacking fault energy (GSFE) surfaces of all slip planes considered in the literature or observed here within the experimental part. By measuring the energy variation in the region of local minima, one can define the preferable shear vector and estimate the energy barrier between neighbouring local energy minima [34]. In the following, the corresponding γ-surfaces will be given. The γ-surfaces considered within this section are calculated using the Liu potential [28], as they show more similar energy landscapes and closer local energy maxima and minima to the DFT calculations by Wang et al. [34]; as compared to the Apostol and Mishin potential [29]. The results obtained with the Apostol and Mishin potential are in qualitative agreement with the calculations using the Liu potential for most, but not all, γ-surfaces and have been included in the Supplementary Materials for reference (see Figure S1-S4).

### GSFE of the {110} plane and the {211} plane

The GSFE´s of the {110} plane (Figure 9 a)) reveals three local minima ($\gamma_1$-$\gamma_3$). One minimum, $\gamma_2$, is located in <001> direction, which is in agreement to the energy landscape obtained by the Apostol and Mishin potential (see S1 a)). The required maximum energies for shear are lower for the Liu potential [28], compared to the Apostol and Mishin potential [29].

The energy landscape of the {211} planes consists of four local minima ($\gamma_1$-$\gamma_4$), see Figure 9 b). One of these minima ($\gamma_4$), is located in ½ <111> direction, similar to the results obtained using the Apostol and Mishin potential [29] (S1 b)) and results obtained by Wang et al. [34].

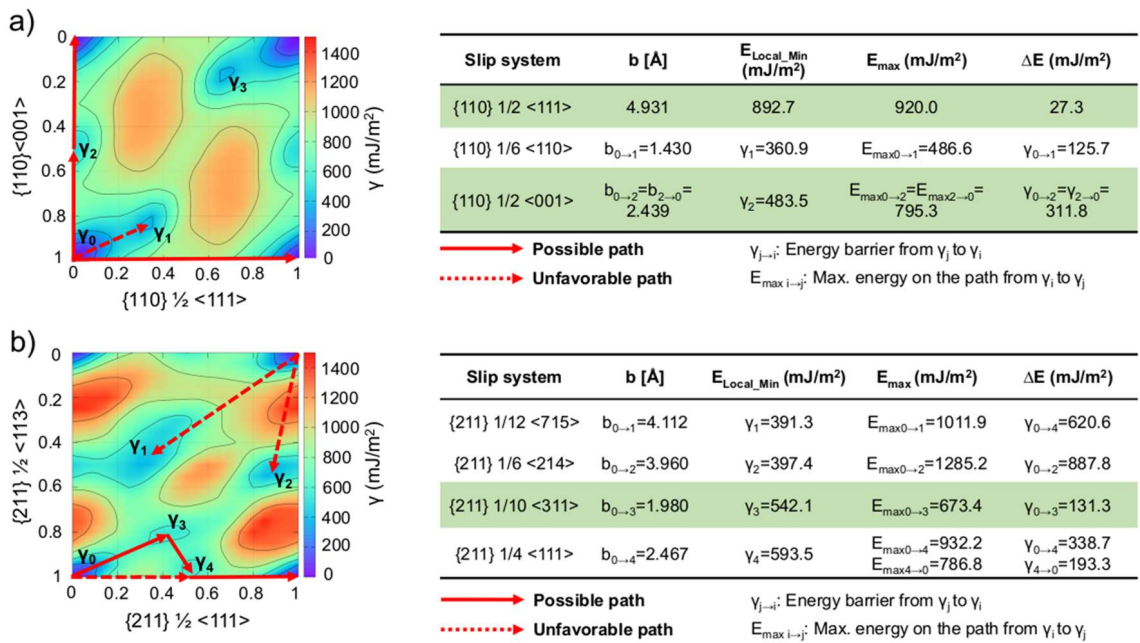

Figure 9: The generalized stacking fault energy surface (γ-surface) of the a) {110} and, b) {211} plane is given including the potential path of slip, marked by the red arrows. Additionally, the length of the corresponding Burgers vectors and the associated energy levels are given and, if considered to represent the actual shift direction, marked in green.

### GSFE of the {002} plane and the {200} plane

The γ-surface of the {002} plane reveals one local minimum in <110> direction (Figure 10 a)). However, as the energy maxima on the potential shear path are, at 1296.5 mJ/m², the highest observed within our simulations, dislocation motion on this slip system is assumed to be less favourable.

With the unstable stacking fault energy on the {200} plane being 1217.6 mJ/m² (Figure 10 b)), the required energies for slip on the {200}<001> and {200}<010> slip system are expected to be of similar magnitude as the ones observed for {002}<110>. Therefore, the probability of dislocation motion on this plane is assumed to be rather small. Similar γ-surfaces were obtained when using the Apostol and Mishin potential [29] (Figure S2).

Due to the relatively high energy barriers on the {200}<001>-, {002}<100>- and {002}<110>- slip systems for both potentials, as well as the consequently low probability of dislocation motion to occur on these slip systems, they are not considered further in the following evaluation.

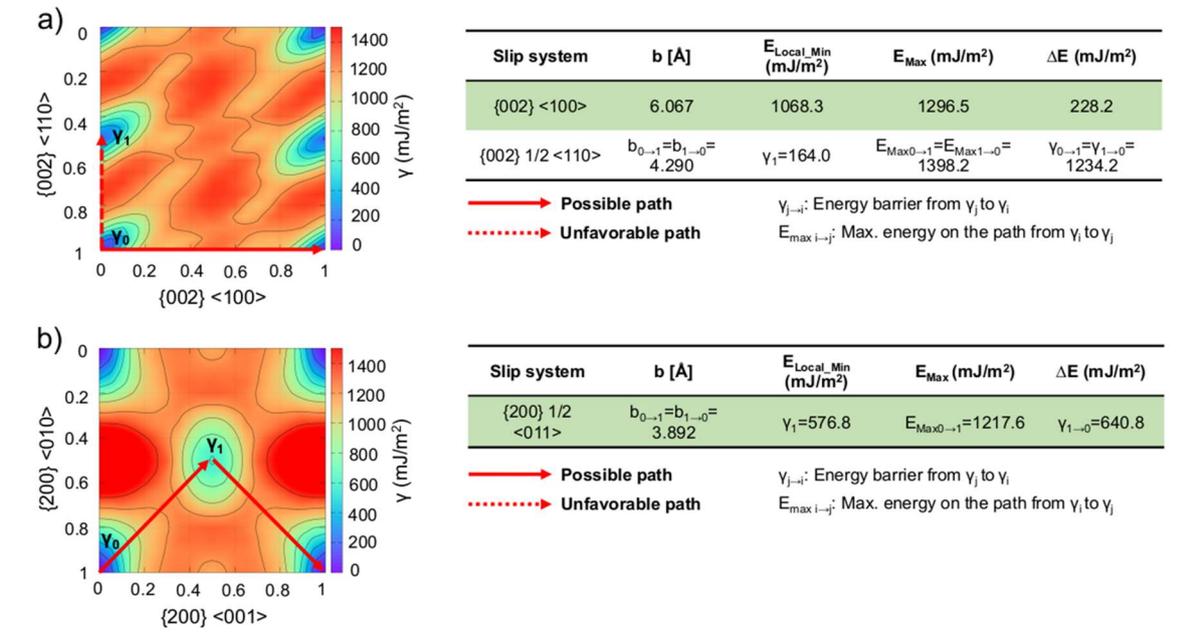

Figure 10: The generalized stacking fault energy surfaces (γ-surface) calculated according the Liu potential [28] of the a) {002} and b) {200} plane are given including the potential path of slip, marked by the red arrows. Additionally, the length of the corresponding Burgers vectors and the associated energy levels are given and, if considered to represent a potential shift direction, marked in green.

GSFE of the {112} plane and the {310} plane

The energy landscape of the {112} plane reveals two minima ($\gamma_1$-$\gamma_2$) with one minimum, $\gamma_1$, located in <110> direction. Shear along this direction requires intermediate maximum energies compared to the other slip systems (Figure 11 a)).

Shear on the {310} plane in <001> direction passes the local minimum $\gamma_1$ (Figure 11 b)) according the GSFE of both potentials. Further, for shear in ½ <131> direction, the full length of the Burgers vector accounts for 9.89 Å, which is relatively large. The latter slip system is therefore not further considered within the analysis.

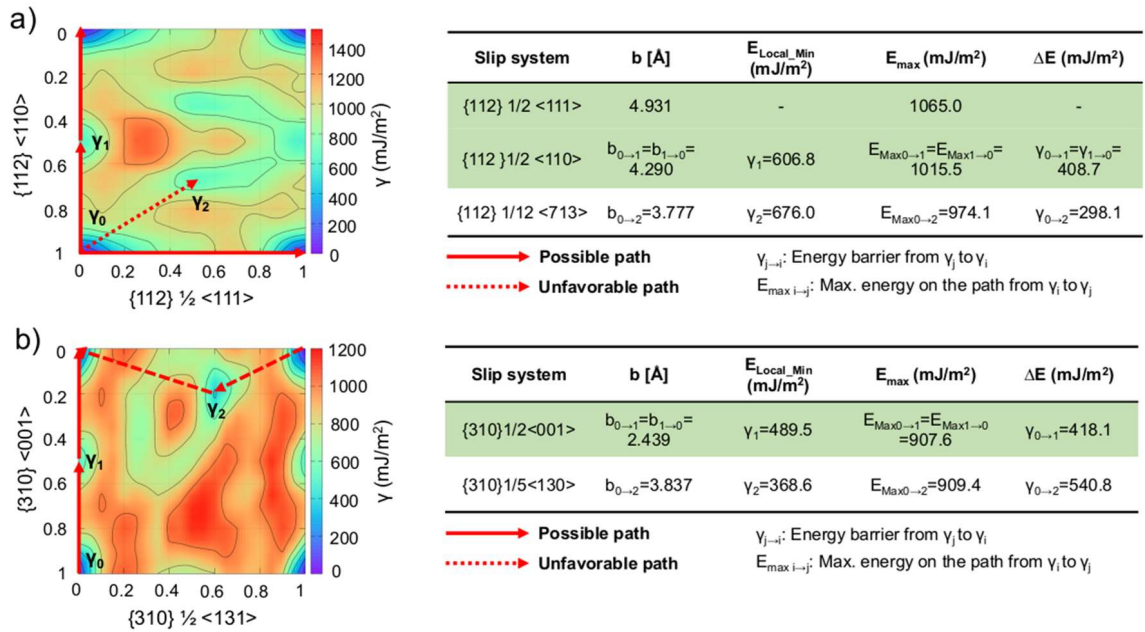

Figure 11: The generalized stacking fault energy surface (γ-surface) of the a) {112} and, b) {310} plane is given including the potential path of slip, marked by the red arrows. Additionally, the length of the corresponding Burgers vectors and the associated energy levels are given and, if considered to represent the actual shift direction, marked in green.

GSFE of the {022} plane

The γ-surface of the {022} plane in ½<$\bar{1}$11> and ½<111> direction reveals three local minima ($\gamma_1$-$\gamma_3$), see Figure 12 a), whereas the {022} plane in <011> and <100> direction reveals four local minima ($\gamma_1$-$\gamma_4$), see Figure 12 b). The required energies for slip in ½<111> and <100> direction are relatively low. A similar energy landscape was obtained when using the Apostol and Mishin potential [29] (see S4, Supplementary) and also using DFT calculations by Wang et al. [34].

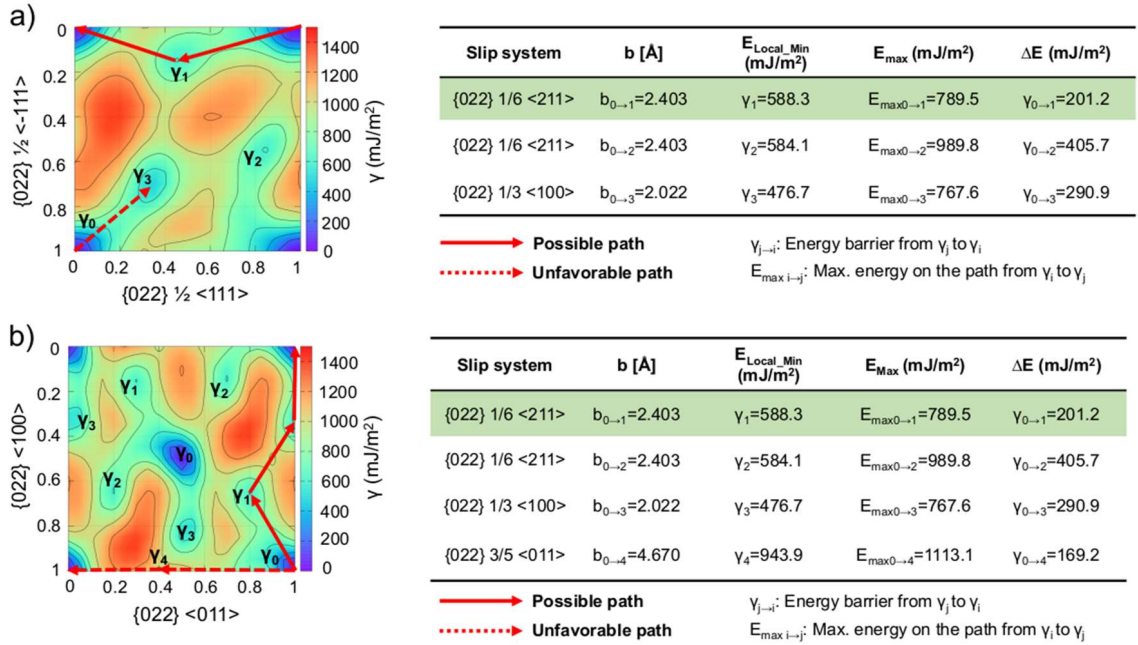

Figure 12: The generalized stacking fault energy surface (γ-surface) of the {022} plane in a) ½<$\bar{1}$11> and ½<111> and b) <100> and <011> direction is given including the potential path of slip, marked by the red arrows. Additionally, the length of the corresponding Burgers vectors and the associated energy levels are given and, if considered to represent the actual shift direction, marked in green.

## 4 Discussion

In the following section, we discuss our experimental and computational results and the geometrical consideration of the interplanar distance, the effective Burgers vector and the GSFE´s for the individual slip systems (Section 4.1) before combining these insights in Section 4.2.

### 4.1 Discussion of individual slip systems

{110} plane

Due to the largest effective interplanar distance between $Al_1$ and $Al_2$ and layers $Al_3$ and $Al_4$ (Figure 5 a)), it is assumed that dislocation motion occurs on an $Al_1$-$Al_2$ shear plane or on an $Al_3$-$Al_4$ shear plane. This is consistent with results from atomistic studies by Zhou et al. [19], where a dislocation with an extended core was found on the $Al_2$-$Al_1$ shear plane and a dislocation with a condensed core on the Cu-$Al_1$ shear plane. According to the γ-surface of the {110} plane (Figure 9 a)), dislocation dissociation is assumed to occur for shear in <001> direction via $γ_2$. The shear vector corresponds to ½<001> and has a length of 2.439 Å. The dissociation of this dislocation is in agreement with atomistic simulations by Zhou et al. [19], our calculations using the Apostol and Mishin potential [29] (see Figure S1 a), Supplementary Material) and experimental observations by Yang et al. [35]. According to Zhou et al. [19], the dissociation width for {110}½<001>-dislocations should be 1.25 nm. Experimental TEM observation of the stacking fault [35] revealed a dissociation width of 10-20 nm. Additionally,

HR-TEM investigations by Bonnet et al. [36] revealed undissociated dislocations on the {110} plane. However, due to experimental restrictions, their edge component was not identified.

The low energy barrier for slip in ½<111> direction of 27.3 mJ/m$^2$ (186.1 mJ/m$^2$ for Apostol and Mishin potential, see Figure S1 a), Supplementary Material) from metastable state to the energy maximum indicates that no stable stacking fault is expected to exist and thus slip via motion of full dislocations should occur. This is further consistent with atomistic simulations using the Apostol and Mishin potential by Zhou et al. [19].

Further, the {110} plane was often reported in literature as slip plane in Al$_2$Cu [12, 14, 16, 19]. Kirsten [14], Ignat et al. [18], Hamar et al. [16] and Zhou et al. [20] reported either slip or cleavage fracture on this plane, indicating easy activation of plastic deformation on this plane. This is furthermore consistent with our findings of micropillar 1-3, where slip on {110}½<111> was observed.

<u>{211} plane</u>

Shear on the {211} plane might occur in the shortest direction, which is ½<111> with a length of 4.931 Å (Figure 6 e)). According to the GSFE of the plane, shift in ½<111> direction is assumed to occur via γ$_3$ (Figure 9 b)). This implies the dissociation of the dislocation into partials, with a b$_{eff}$ of 1.98 Å. This is in agreement with DFT calculations by Wang et al. [34], where the existence of a stable stacking fault on the $\{\bar{1}21\}$ plane was discussed. The dissociation of a dislocation on this plane is further proven by TEM-investigations by Wang et al. [34], where a stacking fault with unknown shear on the {211} plane was observed. Slip on the {211} plane was so far only observed in nanoscaled Al-Al$_2$Cu eutectics [37]. The authors [34] therefore concluded that slip on this plane must be connected to slip continuity through the Al-Al$_2$Cu interface. However, as our experiments reveal slip on {211}½<111> in the pure Al$_2$Cu phase, it is assumed that not only slip through interfaces can facilitate the activation of this slip system, but that it is a generally active system in the Al$_2$Cu phase, as our low CRSS values indicate that slip on this plane may be easy to activate.

<u>{002} and {200} plane</u>

For slip on the {002} plane in <100> direction, slip might occur in any of the layers (Figure 6 b)). For slip in <110> direction (Figure 6 c)), however, it is not clear if slip can occur due to the different interatomic distances in the Al-layer. In the upper Al-layer, vector b$_1$ connects atoms with small interatomic distances followed by larger interatomic distances, whereas vector b$_2$ connects atoms of the same interatomic distance. The assumption that dislocation motion on the {002} and {200} planes is less favourable due to the high required energies is in agreement with only one publication by Ignat et al. [18] reporting the observation of slip on the {002} and {200} plane. However, their analyses were performed after elevated temperature deformation (320°C) in a eutectic specimen and due to both the presence of many interfaces and the much stronger effect of thermal activation, the activation of slip systems that are difficult to activate at low temperature is not surprising and does therefore not necessarily contradict our findings. Additionally, Ignat et al. [15] observed dislocations on the {200}-plane after their elevated temperature creep tests. However, they did not take the effect of ordering of the intermetallic phase into account and might therefore have neglected other potential slip systems during their analysis.

{112} plane

According to the γ-surface of the {112} plane (Figure 11 a)) shear in ½<111> direction might occur via motion of full dislocation. In <110> direction, however, the local minimum $γ_1$ on the shear path might lead to the dissociation of dislocations. Due to the relatively high energy maxima which need to be overcome, the probability for slip in these directions is assumed to be relatively low. However, the {112} plane was observed as active slip plane by Ignat et al. [15, 18], as fracture plane by Hamar et al. [16], and also a spiral dislocation on this plane was reported by Bonnet et al. [38].

{310} plane

The {310} slip plane consists of layers containing Al and Cu atoms. As the interatomic spacing between Al-Al and Cu-Cu is neither the same nor a multiple of that of Cu-Cu (Figure 7 e)), it is not clear how a dislocation might move in <001>-direction. According to the γ-surface, it is further questionable whether dislocation motion on the {310} plane is likely to occur, with maximum required energies of 907.6 and 909.4 mJ/m$^2$ respectively. This is in agreement with the fact that no experimental studies showed slip on {310} planes, only Zhou et al. [19] investigated dislocation glide on the {310}-plane at 600 K and an activation energy of 4.5 GPa using MD simulations. None of the micropillars deformed within this study revealed slip on this slip plane.

{022} plane

So far, slip on {022} planes was observed by several researchers [9, 14, 18], however, due to its low density when taking the effect of ordering not into account, the reason for this observation was not understood or its occurrence considered unusual. According to the γ-surface (Figure 12), slip in ½<111> direction might take place via $γ_1$, which has a relatively low $E_{max}$ of 789.5 mJ/m$^2$ and therefore might occur via partial dislocation motion. Further, the corresponding γ-surface might indicate the formation of a stable stacking fault, as the energy barrier for shear on the {022} plane is larger than the one for the {211} plane, where a stable stacking fault was observed. As Wang et al. [34] doubted the stability of the stacking fault on the {022} plane based upon their DFT calculations, we cannot conclusively decide here whether or not this stacking fault might be stable. We therefore assume that dislocation motion via partial dislocation on this plane should be possible, independent of the stability of the corresponding stacking fault. The lack of experimental observation of this stacking fault points towards the instability of the stacking fault, which is in agreement with the conclusions by Wang et al. [34]. A similar dissociation on the γ-surface of the {022}-plane in ½<111> direction was also revealed by using the Apostol and Mishin potential, see Figure S2 a) in the Supplementary Material.

According to the γ-surface of the {022} plane (Figure 12 b)), slip in <100> direction might also occur via $γ_1$. Therefore, the occurring mechanism should be the same as the one discussed above, where shear occurs via partial dislocations together with an unknown stability of the stacking fault. For slip in <011> direction, however, which might occur via $γ_4$, the maximum energy which needs to be overcome is 1113.1 mJ/m$^2$, which is relatively high. Therefore, the probability of slip to occur in <011> direction is low. Similar results were revealed using the Apostol and Mishin potential, where <100>-slip might occur via $γ_1$, and the maximum energy for <011> slip is also high (1274.3 mJ/m$^2$, see Figure S2 b) in the Supplementary Material).

### 4.2 Combination of effective interplanar distances and effective Burgers vectors

The shear stress needed to move an edge dislocation on a certain slip system (at 0 K) can generally be estimated using the Peierls-Nabarro equation [39, 40], Eq. 2:

$$\tau_P = \frac{2\mu}{1-\nu} \exp\left(-\frac{2\pi d}{(1-\nu)b}\right), \qquad \text{(Eq. 2)}$$

with µ, ν, d and b being the shear modulus, Poisson´s ratio, interplanar distance and Burgers vector, respectively. Hence, the d/b ratio is a good indicator of which slip systems are more easily activated than others. Table 2 summarizes the most dense lattice planes and closest packed directions of the intermetallic $Al_2Cu$ θ-phase including their interplanar distances according to Ignat et al. [15] and calculated d/b values taking into account the lattice parameters defined by Meetsma et al. [33].

Table 2: Slip systems in the $Al_2Cu$ θ-phase including the d/b ratio

| Slip system | Lattice spacing according to Ignat et al. [15] in [Å] | Calculated lattice plane spacing in [Å] according to Meetsma et al. [33] | Calculated length of Burgers vector b in [Å] | d/b ratio, taking into account the lattice spacing after [33] |
|---|---|---|---|---|
| {110} ½ <111> | 4.278 | 4.290 | 4.935 | 0.8693 |
| {110} <001> | 4.278 | 4.290 | 4.877 | 0.8796 |
| {200} <001> | 3.026 | 3.034 | 4.877 | 0.6221 |
| {002} <100> | 2.439 | 2.439 | 6.067 | 0.4020 |
| {211} ½ <111> | 2.366 | 2.371 | 4.935 | 0.4804 |
| {112} ½ <111> | 2.119 | 2.120 | 4.935 | 0.4296 |
| {310} <001> | 1.913 | 1.919 | 4.877 | 0.3935 |
| {022} ½ <111> | 1.900 | 1.901 | 4.935 | 0.3852 |
| {022} <010> | 1.900 | 1.901 | 6.067 | 0.3133 |

However, the slip planes observed in the deformed micropillars (Table 1) are not consistent to the closest packed planes in Table 2. Specifically, the following primary slip systems are unexpected based merely on the d/b ratio: {211}½<111>, {022}½<111> and {022} < 010 >. One of the reasons for the occurrence of these unexpected slip systems might be the complexity of the intermetallic phase, as the calculated interplanar distances solely represent the distances between crystallographic planes in a simple tetragonal unit cell, not considering the effect of ordering. Furthermore, the dissociation of full dislocations into partials would affect the magnitude of the Burgers vector and consequently the Peierls stress.

Therefore, we combined all the information gained in the sections above on effective interplanar distances, effective Burgers vectors and GSFE and determined a new order of slip systems according to the $d_{eff}/b_{eff}$-ratio.

In the following, all of the above-mentioned slip systems with reasonably low maximum energies for shear will be considered according to their calculated $d_{eff}/b_{eff}$-ratio, see Table 3. Note that in Table 3, dissociation of all dislocations is assumed to occur wherever an energy minimum was found along the shear path. This dissociation does not necessarily result in the formation of a stable stacking fault. All experimentally observed slip systems, which are marked in bold, correspond to planes requiring low maximum energies for shear in our simulations.

Table 3: Possible slip systems of the $Al_2Cu$ θ-phase with the manually determined effective interplanar distance, $d_{eff}$, when considering the ordering of the complex unit cell, the magnitude of the effective Burgers vector, $b_{eff}$, calculated using MS simulations and the $d_{eff}/b_{eff}$-ratio. The slip systems which were found within this study are marked in bold and a potential dissociation of the Burgers vector is given. Additionally, the calculated maximum energy $E_{max}$ for shear is given according to the Liu potential [28]. Slip systems corresponding to a high $E_{max}$ (> 1200 mJ/m$^2$) are further marked in italics.

| Slip system | $d_{eff}$ [Å] | $b_{eff}$ [Å] | Full/ partial dislocation | $E_{max}$ [mJ/m$^2$] | $d_{eff}/b_{eff}$ |
|---|---|---|---|---|---|
| **{110} ½ <111>** | 1.3565 | 4.931 | full | 920.0 | 0.2751 |
| {110} <001> | 1.3565 | 2.439 | partial | 795.3 | 0.5562 |
| *{200}<001>* | *1.1151* | *3.892* | *partial* | *1217.6* | *0.2865* |
| *{002} <100>* | *1.2193* | *6.067* | *full* | *1296.5* | *0.2010* |
| *{002} <110>* | *1.2193* | *4.290* | *partial* | *1398.2* | *0.2842* |
| **{211} ½ <111>** | 2.3710 | 1.980 | partial | 673.4 | 1.1975 |
| {112} ½ <111> | 0.7793 | 4.931 | full | 1065.0 | 0.1580 |
| {112} <110> | 0.7793 | 4.290 | partial | 1115.5 | 0.1817 |
| {310} <001> | 1.2133 | 2.4390 | partial | 907.6 | 0.4975 |
| {310}1/2<131> | 1.2133 | 3.837 | partial | 909.4 | 0.3162 |
| **{022} ½ <111>** | 1.9006 | 2.403 | partial | 789.5 | 0.7910 |
| **{022} <100>** | 1.9006 | 2.403 | partial | 789.5 | 0.7910 |
| {022} <011> | 1.9006 | 4.670 | partial | 1113.1 | 0.4070 |

If all slip systems which require a relatively low energy for shear are sorted according to their $d_{eff}/b_{eff}$-ratio (Figure 13), an indication of the activation of slip systems can be obtained. Finally, the slip planes which were observed in our study experimentally (marked in bold) coincide with the slip systems of high $d_{eff}/b_{eff}$-ratio, confirming that the $d_{eff}/b_{eff}$ ratio can give indications for the activation of slip systems not only in elemental metals or semiconductors but also in more complex ordered intermetallic crystals.

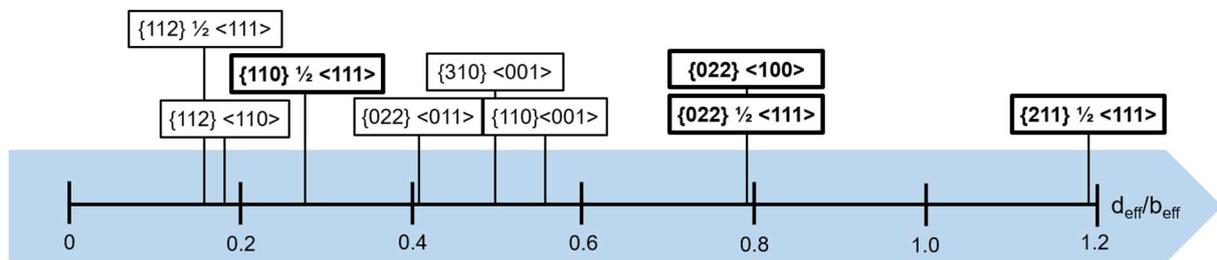

Figure 13: Slip systems of the $Al_2Cu$ phase ordered according to their $d_{eff}/b_{eff}$ ratio as determined within this paper. The experimentally observed slip systems are marked in bold.

The slip systems ordered according to the highest $d_{eff}/b_{eff}$ ratio are: {211}½<111>, {022}½<111> and {022}<100>, {110}<001>, {310}<001>, {022}<011>, {110}½<111>, {112}<110> and {112}½<111> in descending order.

In the following, this order of slip systems (Figure 13) will be compared to the slip systems observed experimentally within this study. Micropillars 1-3 deformed plastically by activating the $(211)½[1\bar{1}\bar{1}]$ slip system. Indeed, micropillars 1-3 have a high Schmid factor for this slip system (0.47) and it is also the slip system with the highest $d_{eff}/b_{eff}$-ratio. The secondary slip system of micropillars 1 and 3 is $(\bar{1}10)½[\bar{1}\bar{1}\bar{1}]$. The Schmid factor of this slip system is 0.48 in micropillars 1-3. As all slip systems with a higher $d_{eff}/b_{eff}$ ratio than the secondary slip system have lower Schmid factors (0.38 for {022}<100> and 0.32 for {022}½<111>, 0.34 for {110}<001>, 0.43 for {310}<001> and 0.33 for {022}<011>), the deformation behaviour of these pillars can be understood using the new approach. For micropillar 4, the $(02\bar{2})½[\bar{1}\bar{1}1]$ and the $(121)½[\bar{1}1\bar{1}]$-slip systems have the same Schmid factor of 0.48. And indeed, slip on both slip systems is observed, in agreement with the former defined order of slip systems.

All of the occurring slip systems of micropillars 5-15 can be explained by considering Schmid´s law together with the determined $d_{eff}/b_{eff}$ ratios.

This assumption that the activation of slip systems in $Al_2Cu$ is related to the $d_{eff}/b_{eff}$ ratio (Figure 13) is also consistent to slip systems reported in literature. The reported "unusual" slip on {022} planes [9], [14], [18] is found not be unusual at all but rather be due to the large $d_{eff}$ when considering the ordering. The reason for the disregard of the {211}½<111>-slip system in most of the former studies might be related on the so far used "conventional" approach to estimate the order of slip systems according to the $d/b$-ratio. According to this, the {211}½<111>-slip system is only fifth in order (see Table 2) and thus not expected to slip. The additional decrease in Burgers vector due to dislocation dissociation resulted in a large $d_{eff}/b_{eff}$ ratio.

Finally, also the experimentally measured CRSS values follow the $d_{eff}/b_{eff}$-ratio as lower measured CRSS coincide with larger $d_{eff}/b_{eff}$ values. Specifically, the CRSS value of {211}½<111> is smallest with 0.76 ± 0.05 GPa, followed by a CRSS value of 0.88 ± 0.07 GPa for {022}½<111> and 0.97 ± 0.12 GPa for {022}<100>. For slip systems with lower $d_{eff}/b_{eff}$ ratios, no data is so far available, as they were not activated as primary slip system.

## 5. Conclusions

We conducted micropillar compression experiments on the $Al_2Cu$ Θ-phase at ambient temperature. The obtained slip systems were analysed using a geometrical consideration of the interplanar distances in conjunction with MS simulations to obtain the GSFE and effective Burgers vectors. The obtained results are:

- Plastic deformation of the $Al_2Cu$ Θ-phase at ambient temperature was observed on the following slip planes and approximate slip directions: {211}½<111>, {110}½<111>, {022}½<111> and {022}<100>.
- As slip on the {211}- and the {022}-plane is considered "unusual", a new order of slip systems was proposed based on the here defined effective interplanar spacing and magnitude of the Burgers vector. This new order is: {211}½<111>, {022}½<111> and {022}<100>, {110}<001>, {310}<001>, {022}<011>, {110}½<111>, {112}<110> and {112}½<111>, from high to low ratio of $d_{eff}/b_{eff}$.
- The critical resolved shear stresses of the observed primary slip systems were measured as 0.76 ± 0.05 GPa for the {211}½<111>-slip system, 0.88 ± 0.07 GPa for the {022}½<111> slip system and 0.97 ± 0.12 GPa for {022}<100>, which is in agreement with the proposed new order of activation of slip systems.

Published in Acta Materialia, Volume 209, 1 May 2021, 116748

## Acknowledgement

The authors gratefully acknowledge funding of the priority program "Manipulation of matter controlled by electric and magnetic field: Towards novel synthesis and processing routes of inorganic materials" (SPP 1959/1) by the German Research Foundation (DFG). This work was supported by grant number 319419837. In addition, this project has received funding from the European Research Council (ERC) under the European Union's Horizon 2020 research and innovation programme (grant agreement No. 852096 FunBlocks). Simulations were performed with computing resources granted by RWTH Aachen University under project (rwth0591). We furthermore thank Dr. J. Guénolé for fruitful discussions.